# New Hashing Algorithm for Use in TCP Reassembly Module of IPS

Sankalp Bagaria
Centre for Development of Advanced Computing
sankalp[at]cdac[dot]in

**Abstract**

Since last decade, IDS/ IPS has gained popularity in protecting large networks. They can employ signature – based techniques and/or flow-based techniques to prevent intrusion from outside/ inside the network they are trying to protect. Signature – based IDS/ IPS can be stateless or stateful. Stateful IDS can store the state of the protocol and use it for better detection of malware. In the case of TCP/IP networks, an attacker can also launch an attack such that the malicious code is distributed over many packets. These packets pass through the traditional IDS/ IPS and reassemble inside the network. Once re-assembled inside the network by the TCP/IP layer, the malicious code launches an attack.

The TCP state and a copy of last few packets for each active connection has to be maintained in IDS/IPS. In TCP re-assembly, packets are re-assembled at IDS/IPS and searched for signature matches. A connection table has to be maintained for active connections and their list of last few (atmost 11) packets that have already arrived. We need data structures for searching the connection that the latest incoming packet belongs to. Popular hashing algorithms like CRC, XOR, summing tuple, taking modulus are inefficient as hash keys are not evenly distributed in hash-key space. Thus we show how an algorithm based on cryptography concepts can be used for efficient hashing in network connection management. We also show how to use full four – tuple for calculating hash key instead of simply summing the tuple and taking the modulus of the sum.

**Introduction**

An intrusion detection/ prevention system (IDS/ IPS) is a security device, usually situated at the periphery of the network it wants to protect. An intrusion detection system (IDS) generates an alert when it sees suspicious packets, while an Intrusion Prevention System (IPS) does active blocking of malicious packets. An IPS/IDS is better than firewall in detecting malicious data. It detects/ prevents malicious code like virus, worm, trojan horse etc from harming the network. It can stop both external and internal threats.

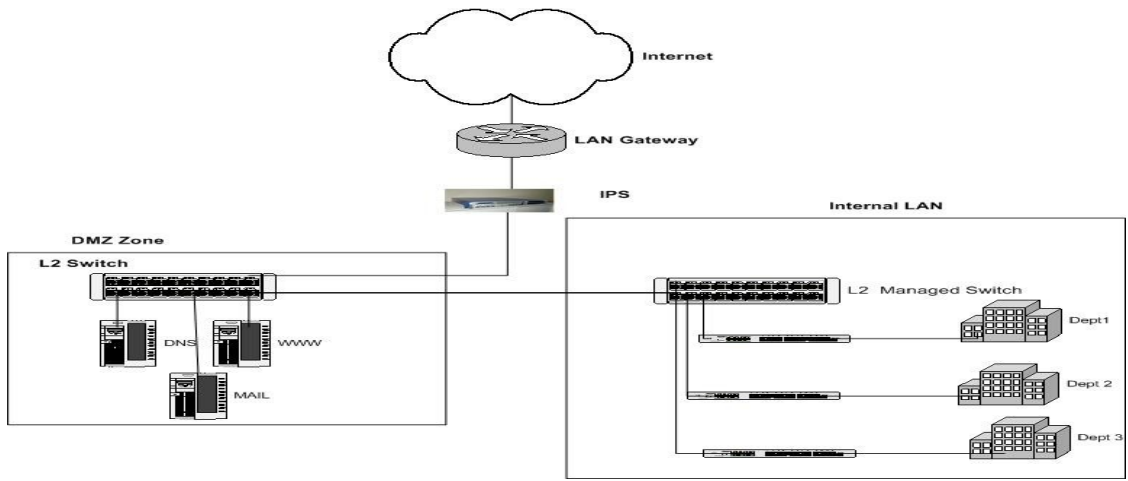

IDS/ IPS employ signature based detection or statistical anomaly based detection or a combination of both. The signature – based detection is very effective in preventing known attacks. Flow based methods look for statistical patterns in network flow that indicate an attack. The flow-based IDS/ IPS look at statistical information instead of particular packet content.

An example of an alert of popular IDS – Snort is as follows:

alert tcp $EXTERNAL_NET ANY -> $HTTP_SERVERS $HTTP_PORTS (msg :"WEB - CGI FINGER ACCESS"; flow : to_server, established; uricontent :"/finger"; nocase; reference : arachnids , 221; classtype : attempted-recon; sid:839; rev:7;)

The signature 839 says that an alert has to be fired if the TCP connection is established with TCP flow from TCP client to TCP server and a packet arrives from any external network IP address to any "HTTP SERVERS" ip addresses, from any TCP port number to any "HTTP PORTS" port number, and if the uricontent of HTTP is "/finger".

In the case of signature based detection, stateful inspection can be employed for better results. Unlike stateless IDS, stateful inspection maintains protocol state for better signature – matching. For example, the first packet of a new connection must be the SYN packet according to the TCP protocol. So, we can drop any new connection that doesnt have SYN flag set in its first packet. Further to simply storing TCP state and matching with appropriate signatures as in the example above, we can also employ TCP re-assembly. If a clever adversary distributes malicious content over several packets instead of sending malware in a single packet, he will succeed in bypassing IDS/ IPS since traditional IDS/IPS look at just one packet at a time. The multiple packets will assemble in target network and attack would be launched.

In TCP re-assembly, along with protocol state, actual copies of packets is stored for matching. When a packet is received by IDS, it is re-assembled with copies of previous packets and then it is matched against signatures. A TCP segment containing malicious code can be divided in atmost 12 packets. Thus, for each connection, we will have to store atmost last 11 packets in IDS/IPS. When a new packet pertaining to an existing connection comes, we add this latest packet to the end of the existing list of packets for that connection, re-assemble them and check for malicious content. An alert is raised if there is a match otherwise the latest packet is added to the tail of the connection's linked list and the first packet is dropped such that the total number of packet-copies stored in IDS for that connection is less than 12.

The assembled segment – data for the connection may also be stored for future use. When the first packet is removed and new packet added to the list, the first part of that connection's segment data may be removed and segment data of new packet be added such that the re-assembly time is decreased whenever next packet comes. This will increase throughput.

We should note that the actual packet is not held till there is a match. Since, even if 11 packets have gone inside the network, an attack can be launched only when the 12th packet comes. Thus the previous 11 actual packets need not be held for any connection. If there is a match, 12th packet is prevented in IPS to go inside the network or an alert is generated in IDS, as the case may be.

We need a data structure – connection table - for storing the copies of the last few packets for all active connections. This data structure should allow for searching for a particular connection in the connection - table and adding/ removing connection from connection table data structure. We would need an efficient hashing technique for fast and efficient retrieval of connection entries. Our paper is about this data structure.

**Related Work**

IDS/ IPS is a very mature technology and lot of work has been done on it. Both stateless (no memory of previous packets) and stateful inspection (maintaining state of protocol) has been explored. Lot of work has been done on connection management. TCP re-assembly is gaining interest for inclusion in IDS/ IPS. Various data structures have been explored. Snort [1] is a popular IDS. [2] is a patent which discusses a solution to handle storing and retrieving TCP connection state information. It is based on hardware implementation such as ASIC and FPGA. [3] is a good background on TCP re-assembly. [4] gives a data structure that can be used in TCP re-assembly. [5] gives an architecture that can be used in TCP re-assembly. [6] is another useful reference for IDS evading techniques and TCP Re-assembly. [7] suggest using bloom filter for hashing problem.

**System architecture**

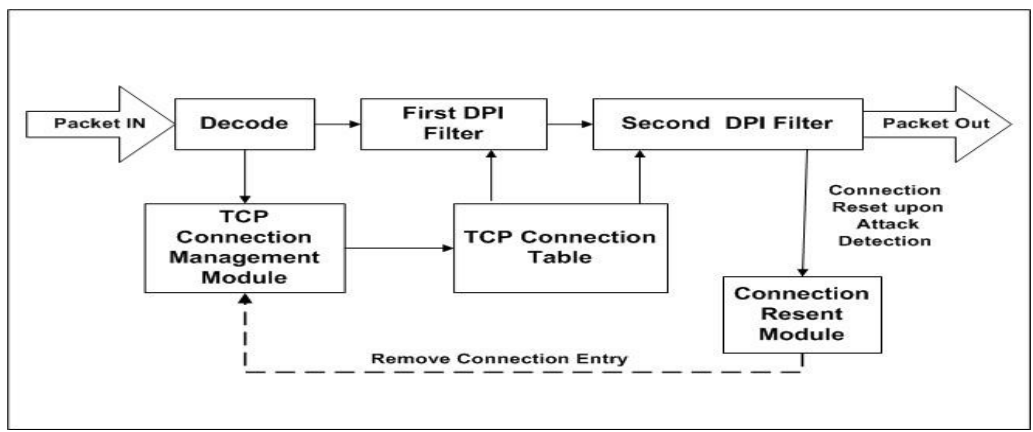

Incoming packet is decoded and passed through first DPI filter where signatures are matched packet-wise. A copy goes to TCP Connection Management module. The TCP connection table is searched and appropriate connection is found. This goes to second DPI filter. In second DPI filter, TCP Re-assembly takes place. And the assembled segment is compared with signatures. If there is a match, connection is reset. A reset signal is sent to connection management module also for updating connection table. A thread for detecting and removing connections that have been inactive for a period of time (say, 30 minutes) from the connection table is also employed. If there is no match, the data is allowed to go to its destination.

**Data Structures and Algorithms involved**

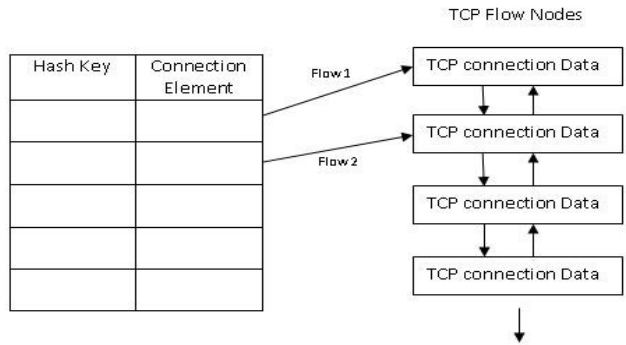

A TCP connection table is prepared to store hash – key and corresponding connection element. The connection table allows addition and removal of TCP Connection Element. Connection Element points to head of the linked list of TCP connection data. Each TCP connection data comprises of one connection - 4 tuple and a linked list consisting of atmost 11 segment-parts. Malicious segment may be distributed over the segment of max 16384 bytes. This data maybe fragmented in ceiling ( 16384 / 1460) = 12 packets (1460 is maximum packet size allowed in network.) Thus we have chosen to store atmost 11 packets (incoming packet being the 12th). These copies of packets are chained in a linked list, and viewed as a queue. We can also store re-assembled segment separately for future addition of new packet – segment and to be used for matching with signatures.

This combination of max 12 packets is re-assembled at IDS/IPS and the segment is compared against signatures looking for malicious data. If there is a match, an alert is generated as in IDS or packet is dropped as in IPS. If there is no match, the latest packet is allowed to go to its destination, and a copy of the new packet is stored at the tail – packet of the connection data. If there were already 11 packets in the list, first packet from the head of the list is dropped. As said before, we can keep the assembled segment in the data structure for future use. We will need to remove the first part of the segment and add the new packet's segment – data.

An important part of this table is hash – key, a mechanism of finding connection, matching the current packet's connection from the table. Hash – key should be such that it distributes connections uniformly over the table and also allows fast searching for appropriate connection in the connection table.

A common method to finding hash – key is to calculate sum of 4 – tuple (source IP address, destination IP, source port and destination port). This sum mod N gives the hash – key (N being table size). Collisions are resolved by adding new connection data to a chained linked list of connection data (head of the list is pointed by connection element). As can be easily seen, this method is flawed because it uses only the sum of 4 – tuple and not the actual 4 – tuple. Since IP addresses and port numbers are not uniformly distributed in their entire range - space, the summation followed by taking mdulus of the sum of tuple causes more collisions and thus, under - utilization of the hash-key space. Also time is wasted in resolving collisions at run-time.

In this paper, we suggest use of the cryptographic algorithm for calculating hash – key of 4 – tuple and make sure that full connection table is used. We append 4-tuple one after another ie, Source IP, Destination IP, Source Port, Destination port to get 96 bit number. This is hashed using one of the symmetric key block cipher or one of the cryptographic hashing algorithms to get a number of size same as the connection – table (chosen as a power of 2 for convenience). The cryptographic key can be fixed to an appropriate random number for entire duration.  As symmetric key algorithms hash a number (here, 96-bit 4-tuple) to random space of size N (here, N  is size of connection table - chosen as a power of 2 for convenience) uniformly, whole connection table is used.

If we use the same key throughout, we get a uniform mapping of 4-tuple to connection – table consisting of max N connection entries. If we use the same key to map incoming packet's 96-bit 4-tuple, we get a number that gives the appropriate connection entry in table of size max N connection entries. We also see this method maps incoming and outgoing packets to separate hash - keys. In traditional method of calculating hash – key done by summing the 4-tuple and taking modulus, the incoming and outgoing packet map to same location. This is resolved by comparing incoming 4 - tuple source IP, destination IP, source port and destination port with that of the connection (signifying direction of flow) in the connection table in the second step. But our method resolves the incoming and outgoing packets automatically by mapping them to different locations in the connection – table. As our method reduces collisions by using full connection – table, time required to resolve collisions at run-time is prevented.

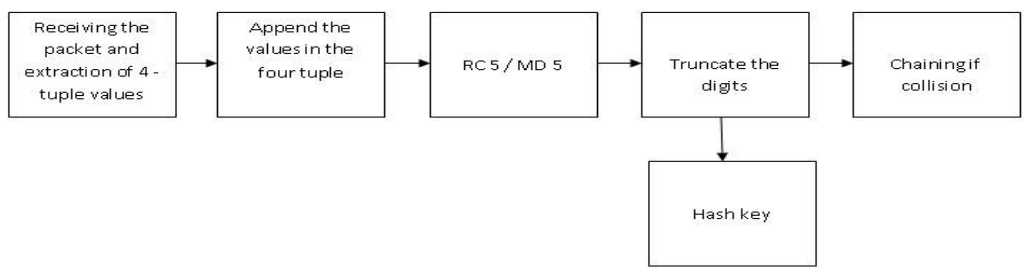

**Acknowledgements**

We are grateful to CDAC for providing us this opportunity to work on the hashing problem. We are also grateful to the authors who have worked on TCP reassembly problem before us.

**Conclusion**

The authors discuss prevalent technology of TCP Connection Management module of IDS/ IPS with special reference to TCP Re-assembly. Then the authors discuss how hashing is important for an efficient way of TCP connection management. Then they discuss prevalent hashing mechanisms. The authors then propose new hashing algorithm. They show how symmetric key block cipher like AES with fixed key, or cryptographic hashing algorithm like MD5 can be used for providing hash-key uniformly distributed over connection table. They also show how full 4 – tuple can be used for efficient hashing instead of using just the sum of 4-tuple.